# Towards monitoring conformational changes of the GPCR neurotensin receptor 1 by single-molecule FRET

Thomas Heitkamp[a], Reinhard Grisshammer[b], Michael Börsch[a,*]

[a] Single-Molecule Microscopy Group, Jena University Hospital, Friedrich Schiller University Jena, Nonnenplan 2 - 4, 07743 Jena, Germany
[b] Laboratory of Cell Biology, National Cancer Institute, National Institutes of Health, 50 South Drive, Bethesda, MD 20814, USA

## ABSTRACT

Neurotensin receptor 1 (NTSR1) is a G protein-coupled receptor that is important for signaling in the brain and the gut. Its agonist ligand neurotensin (NTS), a 13-amino-acid peptide, binds with nanomolar affinity from the extracellular side to NTSR1 and induces conformational changes that trigger intracellular signaling processes. Our goal is to monitor the conformational dynamics of single fluorescently labeled NTSR1. For this, we fused the fluorescent protein mNeonGreen to the C terminus of NTSR1, purified the receptor fusion protein from *E. coli* membranes, and reconstituted NTSR1 into liposomes with *E. coli* polar lipids. Using single-molecule anisotropy measurements, NTSR1 was found to be monomeric in liposomes, with a small fraction being dimeric and oligomeric, showing homoFRET. Similar results were obtained for NTSR1 in detergent solution. Furthermore, we demonstrated agonist binding to NTSR1 by time-resolved single-molecule Förster resonance energy transfer (smFRET), using neurotensin labeled with the fluorophore ATTO594.

## 1. INTRODUCTION

G protein-coupled receptors (GPCRs) are membrane proteins that respond to extracellular signals triggering a variety of intracellular signaling responses. This large superfamily of GPCRs is involved in fundamental physiological processes of neurotransmission, immune response, behavioral regulation, vision, smell and taste [1]. Accordingly, GPCRs are the targets for as much as one-third of all therapeutic drugs today. Understanding the detailed molecular mechanism of how GPCRs function is a vibrant area of research.

We focus on the neurotensin receptor 1 (NTSR1) [2] that belongs to the β group of class A GPCRs. The biochemical protocols to purify this receptor have been developed and optimized [3-5] and several X-ray crystal structures of NTSR1 have been solved [6-9]. Using single-molecule fluorescence methods we aim at unravelling the conformational dynamics of NTSR1 with respect to ligand binding, and G protein interaction, and how a membrane potential possibly influences the receptor activity. We have previously built confocal microscopes [10, 11] to observe individual membrane proteins one-after-another by single-molecule Förster resonance energy transfer (smFRET). Our current study extends smFRET measurements to the NTSR1 system.

Previously, it was shown that NTSR1 is monomeric in detergent solution but dimeric at high concentrations after reconstitution in a 'native' brain polar lipid environment [12, 13]. Here, we pursue the analysis of the conformational dynamics of monomeric liposome-reconstituted NTSR1 by smFRET. Towards this goal, we fused the brightest green fluorescent protein, mNeonGreen [14, 15], to the C terminus of NTSR1 and measured single-molecule anisotropy of the detergent-solubilized protein and of the receptor reconstituted in liposomes. Counting NTSR1 one-after-another revealed that most of the detergent-solubilized as well as the reconstituted NTSR1 is monomeric. However, single-molecule homoFRET data also indicated minute quantities of dimers under both conditions. A small fraction of very large and highly fluorescent protein aggregates may cause a significant FRET signal in ensemble-averaging fluorescence correlation spectroscopy (FCS) and cuvette experiments.

--------

* email: michael.boersch@med.uni-jena.de

## 2. EXPERIMENTAL PROCEDURES

The rat neurotensin receptor 1 (NTSR1) was genetically modified, heterologously expressed in *Escherichia coli* and purified from the cytoplasmic membrane. The experimental procedures are given below.

### 2.1 Construction of plasmids pNTSR1_mNG and pET15b_mNG5mNG

The pBR322-derived expression vector pRG/III-hs-MBP-T43NTSR1-GFP/reporter-H10 is a pRG/III-hs-MBP derivative [3] and was used as the target vector for the NTSR1-mNeonGreen fusion construct. This template vector harbours the N-terminally truncated NTSR1 gene fused at its 5'-end to the gene encoding the maltose binding protein (MBP) with its signal peptide, which is needed for the correct targeting of the neurotensin receptor to the cytoplasmic membrane of *E. coli* [16]. The construct ends with a decahistidine tag (H10) for purification. The gene encoding mNeonGreen (mNG) was amplified using the primer pair 5'-GCTGATAA**GCGGCCGC**AGTGAGCAAGGGCG-3' and 5'-GCAT**GGTACC**CTTGTACAGCTCGTCC-3' with pACWU-BH1 as the template [15]. The deoxynucleotides corresponding to the mNG gene are underlined and those of restriction sites are in bold letters. The resulting PCR product was restricted with *NotI* and *KpnI* enzymes and then ligated into the correspondingly restricted pRG/III-hs-MBP-T43NTSR1-GFP/reporter-H10 plasmid yielding the vector pNTSR1_mNG encoding an MBP-T43NTSR1-mNG-H10 fusion protein. The construct was verified by DNA sequencing. We refer to the MBP-T43NTSR1-mNG-H10 fusion protein as NTSR1-mNG.

To generate a mNG dimer, the gene encoding the monomer was amplified from pACWU-BH1 thereby introducing a *NdeI* restriction site at the 5'-end and a *BamHI* restriction site at the 3'-end of the mNG gene. After restriction digest, the PCR product was ligated into the correspondingly restricted pET15b vector (Merck, Germany). To fuse a second mNG, a *KpnI* restriction site was introduced into the mNG gene before the stop codon yielding pET15b_mNG_*KpnI*. Then, the unmodified mNG gene lacking the start codon was amplified with a linker sequence at the 5'-end harbouring an *KpnI* site, and with a *BamHI* restriction site directly after the stop codon. The PCR product was restricted with *KpnI* and *BamHI* and ligated into correspondingly restricted pET15b_mNG_*KpnI*. The resulting vector pET15_mNG5mNG encodes an mNG dimer consisting of Gly-Ser-Ser, a hexahistidine tag, followed by Ser-Ser-Gly, a thrombin recognition site and a histidine residue at the N-terminus of the first mNG, followed by the linker Gly-Thr-Gly-Ala-Ser and a second mNG. Constructs were verified by DNA sequencing.

### 2.2 Expression strain and cell growth

The pNTSR1_mNG plasmid was used to transform the *E. coli* DH5α strain ($F^-$ *endA1 glnV44 thi-1 recA1 relA1 gyrA96 deoR nupG purB20 φ80dlacZΔM15 Δ(lacZYA-argF)U169, hsdR17($r_K^- m_K^+$), λ$^-$*). Cells were grown in a 10 l FerMac 320 fermenter (Electrolab, UK) at 37°C, 800 rpm stirring and 6 l/min fresh air using 2xYT medium supplemented with glucose and ampicillin (10 g/l yeast extract, 16 g/l trypton, 5 g/l NaCl, 0.2 % (w/v) glucose, 0.07 g/l ampicillin). Expression of the fusion construct was induced at $OD_{600nm}$ ~ 0.5 by adding 0.25 mM isopropyl-β-D-thiogalactopyranoside. Simultaneously, the growth temperature was reduced to 21°C. The cells were harvested 41 hours later by centrifugation in a Sorvall Evolution RC centrifuge (Thermo Fisher Scientific, USA) at 10,000 x g and 4°C for 6 min. The cell pellets were flash-frozen in liquid nitrogen and subsequently stored at -80°C.

*E. coli* BL21 (DE3) ($F^-$ *ompT gal dcm lon hsdS$_B$($r_B^- m_B^-$) λ(DE3 [lacI lacUV5-T7p07 ind1 sam7 nin5]) [malB$^+$]$_{K-12}$(λ$^S$)*) was transformed with pET15b_mNG5mNG. Cells were grown, induced and harvested as described for the NTSR1 construct with the exception that cells were grown in 2 l quantities in culture flasks with vigorous shaking.

### 2.3 Purification of NTSR1-mNG

The fusion construct was purified as described [3, 5, 17] with minor modifications of the protocol. Unless otherwise noted, all purification steps and pH adjustments were done at 4°C. All chemicals were of highest purity from Sigma-Aldrich or Roth unless otherwise noted. Phenylmethylsulfonylfluoride (PMSF) was always added freshly directly before use.

The frozen cells from one fermenter run (50 g) were crushed between plastic sheets with a hammer. The resulting small pieces were resuspended with a small household blender together with 100 ml of 100 mM Tris/HCl pH 7.4, 60 % (v/v) glycerol (fully synthetic; Carl Roth, Germany) and 400 mM NaCl. While stirring, 0.2 ml of 70 mg/ml PMSF dissolved in ethanol, two complete EDTA-free protease inhibitor tablets (Roche, Switzerland), 1 ml of 1 M $MgCl_2$ solution and 0.15 ml of 10 mg/ml DNaseI solution was added. The cells were lysed and the membranes were solubilized by adding dropwise and under stirring 20 ml of a solution containing 1.2 % (w/v) cholesteryl hemisuccinate Tris salt (CHS) (Anatrace, USA) dissolved in 6 % (w/v) 3-[(3-cholamidopropyl)-dimethylammonio]-1-propane sulfonate (CHAPS) (Glycon, Germany) followed by the dropwise addition of 20 ml of 10 % (w/v) n-dodecyl-β-D-maltoside (DDM) (Glycon, Germany). After stirring for 15 minutes, the receptor solubilization efficiency was enhanced by sonication five times for 1 minute in an ice water bath using a Digital Sonifier 250D (Branson Ultrasonic SA, Switzerland) with the following settings: ½ inch flat tip, 40 % amplitude, 1 s on, 2 s off. The volume of the solubilized material was adjusted to 200 ml, and an additional 0.2 ml of 70 mg/ml PMSF together with one complete EDTA-free protease inhibitor tablet was added and the solution was stirred for a further 30 min. Unsolubilized membranes and cell debris were removed by ultracentrifugation at 257,000 x g for 2 h. The supernatant containing the solubilized receptors was adjusted to 50 mM imidazole and subsequently filtered through a 0.22 µm filter.

All subsequent chromatography steps were done in a cold room using a Pharmacia LKB FPLC (Pharmacia, Germany) modified with the MCR-4V multichannel data logging and visualization system (T&D, Japan). Thereby, it was possible to simultaneously collect UV absorption, conductivity, mixing ratio of the two pumps and fraction pulses from the fraction collector. A custom-made pulse spreader was used to extend the short fraction collector pulses to 500 ms.

All steps of the nickel nitrilotriacetic acid (NiNTA) purification were done with a flow rate of 0.4 ml/min. The supernatant was loaded onto a XK26/20 column (GE-Healthcare, USA) packed with 20 ml NiNTA Superflow resin (Quiagen, Germany) and pre-equilibrated with 50 mM Tris/HCl pH 7.4, 30 % (v/v) glycerol, 200 mM NaCl, 0.5 % (w/v) CHAPS, 0.1 % (w/v) CHS, 0.1 % (w/v) DDM and 50 mM imidazole (NiA buffer). After washing with 200 ml (10 column volumes) of the same buffer, the bound protein was eluted from the column with 60 ml (3 column volumes) of NiB buffer, which is NiA buffer but contains 200 mM imidazole.

The peptide neurotensin (NTS) is the agonist of NTSR1 and binds to a well-defined binding pocket with nanomolar affinity. Therefore, immobilized neurotensin was used for further purification. The NTS resin was prepared as described [3, 18]. In brief, 5 ml tetrameric avidin agarose (Thermo Scientific, USA) was used as resin. However, the high isoelectric point (IP) of avidin would result in high unspecific binding of *E. coli* proteins. Therefore, to decrease the IP, avidin was succinylated by incubation with 2.5 mM succinic anhydride dissolved in 50 mM borate buffer pH 9 for 1 hour at room temperature with constant shaking. Then, the resin was washed with 50 volumes of 50 mM Tris/HCl pH 7.4, 1 mM EDTA followed by the addition of biotinylated NTS (biotin-*β*A-*β*A-QLYENKPRRPYIL) (PanaTecs, Germany) in a twofold molar excess based on the biotin binding capacity of the resin. After shaking 1 hour at room temperature, the resin was packed into a XK16/20 column (GE-Healthcare, USA) and washed with 10 column volumes (CV) of water, followed by 20 CV of 50 mM Tris/HCl pH 7.4, 1 mM EDTA, 5 mM biotin to block free biotin binding sites, followed by 50 CV of 50 mM Tris/HCl pH 7.4, 1 mM EDTA, 0.5 M NaCl, and finally with 10 CV of water. The column was then stored in 50 mM Tris/HCl pH 7.4, 1 mM EDTA, 3 mM sodium azide.

The NiNTA eluate was diluted to 70 mM NaCl with 50 mM Tris/HCl pH 7.4, 30 % (v/v) glycerol, 1 mM EDTA, 0.5 % (w/v) CHAPS, 0.1 % (w/v) CHS and 0.1 % (w/v) DDM (NT0 buffer), supplemented with 0.5 mM PMSF and filtered through a 0.22 µm filter. The NTS column was equilibrated with NT70 buffer (NT0 buffer plus 70 mM NaCl) and afterwards the diluted NiNTA eluate was loaded with a flow rate of 0.4 ml/min. After washing the column with 8 CV of NT70 buffer at a flow rate of 0.5 ml/min, the receptors were eluted with 4 CV of NT1K buffer (NT0 buffer plus 1 M NaCl).

The NTS column eluate was concentrated using an Amicon Ultra 15 (Merck, Germany) ultrafiltration device with a molecular weight cutoff of 50 kDa. Aggregated protein was separated by centrifugation at 25,000 x g for 10 min at 4°C. Subsequently, the supernatant was loaded onto a Superdex 200 10/300 GL size exclusion column (GE-Healthcare, USA) equilibrated with 2 CV of 50 mM Tris/HCl pH 7.4, 15 % (v/v) glycerol, 1 mM EDTA, 0.5 % (w/v) CHAPS, 0.1 % (w/v) CHS, 0.1 % (w/v) DDM, 100 mM NaCl (SEC buffer). The proteins were eluted at a flow rate of 0.3 ml/min and collected in 0.5 ml fractions. The nearly monodisperse peak fractions, containing NTSR1-mNG, were pooled, concentrated *via* a Vivaspin Turbo 4 (Sartorius, Germany) ultrafiltration device (cutoff: 50 kDa), flash-frozen as 5 µl aliquots in liquid nitrogen and stored at -80°C.

### 2.4 Reconstitution of NTSR1-mNG

The NTSR1 fusion protein was reconstituted into preformed liposomes by slightly modifying established protocols [19]. All following steps were done in the dark. Furthermore, all chemicals were of highest purity to minimize fluorescent impurities. All buffers were treated with active charcoal to minimize the fluorescence background. Preformed liposomes were generated by dissolving 100 mg *E. coli* polar lipid extract (Campro Scientific, Germany) in 5 ml chloroform (Uvasol-grade; Merck, Germany). The mixture was evaporated using a rotary evaporator (Büchi, Switzerland) with a water bath temperature of 40°C and additionally dried overnight at room temperature under 20 mbar. Subsequently, the lipid film was resuspended in 10 ml of 50 mM Tris/HCl pH 7.4, 200 mM NaCl, 1 mM EDTA yielding a lipid concentration of 10 mg/ml. The flask was closed under nitrogen gas and was shaken for 1 hour at room temperature until all clumps had dissolved. The preformed liposomes were aliquoted, flash-frozen in liquid nitrogen and stored at -80°C.

The fluorescent fusion construct NTSR1-mNG was reconstituted into preformed liposomes as follows. Preformed liposomes were thawed at room temperature and pelleted by ultracentrifugation at 400,000 x g for 1 hour at 4°C. The pellet was suspended in 50 mM Tris/HCl pH 7.4, 1 mM EDTA (reconstitution buffer) to a lipid concentration of 10 mg/ml, and then four times flash-frozen in liquid $N_2$ and thawed slowly at room temperature. Finally, the vesicles were sized by 11 extrusions through a polycarbonate membrane, first with 1000 nm and then with 100 nm pore sizes (Avestin, Germany). In a 2 ml reaction tube, 100 µl of preformed, sized liposomes and 93.4 µl of reconstitution buffer were mixed with 4.6 µl 10 % DDM under vigorous shaking on a vortexer. The amount of DDM corresponds to a concentration of 4.5 mM, which was determined to be the DDM saturation point $R_{sat}$ of the liposomes under these conditions. After 3 h incubation at room temperature with gentle shaking, either 2 µl of a 1 µM or 10 µM NTSR1-mNG solution in SEC buffer was added, yielding protein concentrations of 10 nM and 100 nM, respectively, which correspond to liposome to protein ratios of 7.5:1 and 0.75:1. After incubation for 1 h at 4° C and gentle shaking, 24 mg of pretreated BioBeads SM-2 (Biorad, USA) were added to remove the detergent. After gentle shaking overnight at 4°C, the proteoliposomes were separated from the BioBeads and used directly for single-molecule measurements.

### 2.5 Purification of the dimeric mNG5mNG

Unless otherwise noted, all steps were done at 4° C. Cells were resuspended in 20 mM sodium phosphate buffer pH 7.5, 500 mM NaCl, 5 mM $MgCl_2$, 20 mM imidazole, 10 % (v/v) glycerol, 0.25 mM PMSF (lysis buffer), supplemented with 1 tablet complete EDTA-free (Roche, Switzerland) and 1 spatula tip DNaseI and RNaseA, lysed by two passages through a cell homogenizer (PandaPlus 2000, GEA Niro Soavi, Italy) at 1000 bar and subsequently centrifuged for 1 h at 50,000 x g. The supernatant was loaded on a 5 ml Ni-Sepharose HP column (GE Healthcare, USA), equilibrated with 20 mM sodium phosphate buffer pH 7.5, 500 mM NaCl, 20 mM imidazole, 10 % (v/v) glycerol, 0.25 mM PMSF (NiNTA buffer). The column was washed with 54 mM imidazole in NiNTA buffer and the mNG dimer was eluted with NiNTA buffer containing 260 mM imidazole. The eluate was concentrated with an Amicon Ultra 15 (Merck, Germany) ultrafiltration device (10 kDa cutoff) and loaded on a self-packed XK26/40 Sephacryl S300 size exclusion column (GE-Healthcare, USA) equilibrated with 20 mM sodium phosphate buffer pH 7.5, 150 mM NaCl, 10 % (v/v) glycerol, 0.25 mM PMSF. The eluted protein was concentrated *via* an Amicon Ultra 15 (Merck, Germany) ultrafiltration device (10 kDa cutoff), aliquoted, flash-frozen in liquid nitrogen and stored at -80°C.

### 2.6 Other biochemical methods

The fluorescently labeled ATTO594-NTS (ATTO594-ELYENKPRRPYIL) was obtained by solid-phase peptide synthesis (PanaTecs, Germany). ATTO647N-CKPRRPYIL was synthesized by stoichiometrically reacting the maleimide of the fluorophore with the truncated derivative of NTS (peptide with a cysteine at its N-terminus, PanaTecs) and used without further purification. Concentrations of NTSR1-mNG and ATTO594-NTS were determined using a Lambda 650 UV/Vis spectrometer (PerkinElmer, USA) and a molar extinction coefficient of 116,000 L·mol$^{-1}$·cm$^{-1}$ for mNG (at 506 nm) and 120,000 L·mol$^{-1}$·cm$^{-1}$ for ATTO594 (at 605 nm). Sodium dodecyl sulfate polyacrylamide gel electrophoresis (SDS-PAGE) was carried out at an acrylamide concentration of 12 %. In-gel fluorescence of mNG was visualized using the G:Box Bioimaging system (Synoptics Limited, UK). Afterwards gels were stained with coomassie-R250.

### 2.7 Custom-built confocal microscope for smFRET and single-molecule anisotropy (homoFRET)

For time-resolved single-molecule anisotropy and smFRET measurements in solution (i.e. homoFRET), a custom-designed confocal microscope was used based on an Olympus IX71 [10, 20-25]. Ps-pulsed linearly polarized excitation was provided with 488 nm at 80 MHz (PicoTa 490, Picoquant) using a 60x water immersion objective (N.A. 1.2). A 150 µm pinhole was employed to reject out-of-focus background. Fluorescence photons were detected by three single-photon counting avalanche photodiodes. Two of them were time-resolution optimized versions (SPCM-AQRH-14-TR, Excelitas) and were synchronized with a ps delay generator with 10 ps resolution (PSD, MPD) for time-resolved anisotropy. Fluorescence of mNeonGreen was detected after passing a 535/70 nm bandpass filter (AHF) and split by a polarization beam splitter for anisotropy measurements. A 580 nm dichroic filter separated NeonGreen from ATTO647N signals (which passed a LP594 longpass filter). Three synchronized TCSPC cards (SPC 154, Becker&Hickl) recorded the photons for simultaneous spectrally-resolved FLIM or anisotropy measurements. Fluorescence time traces were analyzed using the Burst_Analyzer software (Becker&Hickl).

Structured illumination microscopy was performed on a Nikon N-SIM / N-STORM microscope using 488 nm or 647 nm excitation and the 60x water immersion objective with N.A. 1.27 [26, 27]. The 1.5-fold magnification lens was combined with the 2.5-fold magnification lens for the 3D-SIM mode. Images were recorded by a cooled Andor EMCCD camera (iXon DU-897) at 23° C. Fluorescence was measured in the spectral range from 500 to 550 nm for mNeonGreen and from 675 to 725 nm for ATTO647N. Nikon analysis software was used for SIM image reconstruction and for 3D visualization using the maximum intensity projection option. *E. coli* cells expressing NTSR1-mNG were incubated with 100 nM ATTO647N-NTS in PBS buffer for 15 min at 37°C. After centrifugation for 1 min, the supernatant was removed. The cells were resuspended in 2 ml PBS (pre-warmed at 37°C) and shaken for a few seconds at 21°C. The wash step was repeated one more time. 10 µl of the cell suspension were placed on a polylysine-coated glass slide and a cover glass was placed on top to seal the sample chamber. SIM images were recorded immediately.

## 3. RESULTS

### 3.1 The NTSR1-mNG construct

To express functional NTSR1 in *E. coli* the maltose binding protein (MBP) with its own signal peptide was fused to the truncated N-terminus of the receptor (T43NTSR1) to facilitate the insertion into the cytoplasmic membrane of *E. coli* [16]. The first 42 amino acids of NTSR1 were deleted because they harbor protease-sensitive sites; this has no effect on agonist binding. The bright fluorescent protein mNeonGreen (mNG) was fused to the C-terminus of NTSR1 followed by decahistidine tag (H10).

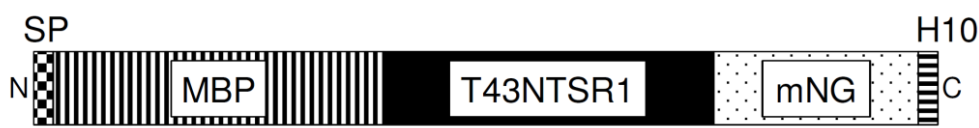

**Figure 1.** The NTSR1-mNG construct. The N-terminally truncated T43NTSR1 is fused to maltose binding protein (MBP) at the N-terminus to aid incorporation into the cytoplasmic membrane, and to a mNeonGreen (mNG) fluorescent protein at the C-terminus. SP refers to the MBP signal peptide and H10 to the decahistidine tag for purification.

### 3.2 Imaging NTSR1-mNG and ATTO647N-NTS in living *E. coli* cells

First, we imaged the fluorescent receptor NTSR1-mNG in living *E. coli* cells to assess protein expression and distribution in the cytoplasmic membrane. Structured illumination microscopy allows imaging of small subjects with a 2-fold improved optical resolution compared to the Abbé limit [28]. The *E. coli* cells as shown in Fig. 2 were elongated and the mNG fluorescence was attributed to the membrane (Fig. 2 A, D). The few bright spots may possibly indicate aggregated receptors. After addition of fluorescent water-soluble ATTO647N-labeled NTS (suitable for excitation with our 647 nm laser of the SIM microscope) to the cells, the ligand was assigned to the membrane of *E. coli* (Fig. 2 B, E). Overlay of the false-colored green mNG and the red ATTO647N images supported a colocalization of receptor and ligand (Fig. 2 C, F), for example as seen for the bacterial cells numbered 1, 2 and 3.

However, binding of ATTO647N-NTS appeared heterogeneous, and some cells did not exhibit significant fluorescence from the ligand (for example, cell 4 in Fig. 2). Note that colocalization imaging by SIM does not provide direct evidence for functional binding of the fluorescent ligand to the receptor because the achievable optical resolution of SIM is in the range of 120 to 150 nm. Direct molecular interaction has to be revealed by FRET imaging *in vivo* or by smFRET measurements *in vitro* (as shown in chapter 3.5 below), respectively.

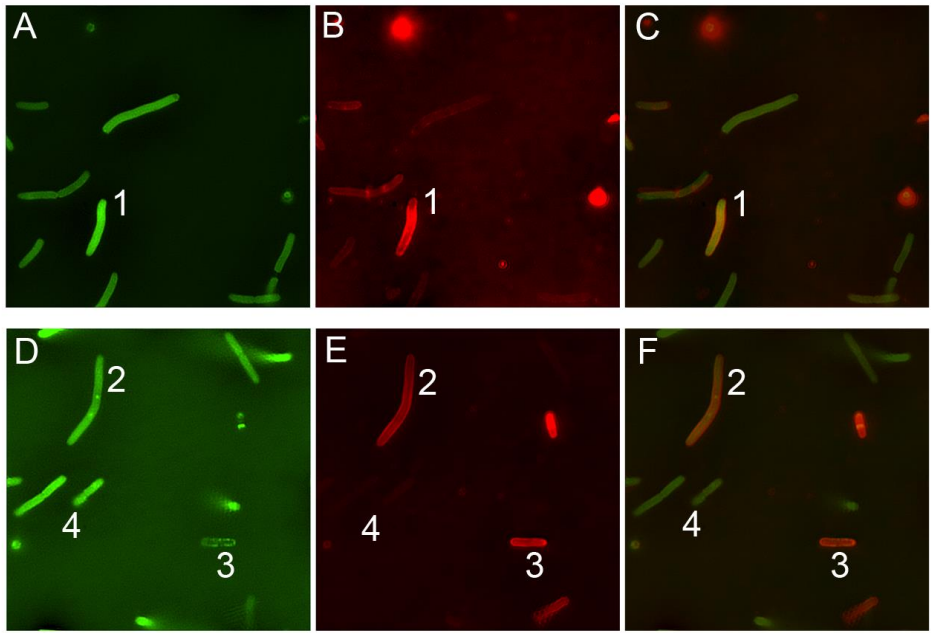

**Figure 2. Colocalization imaging of NTSR1-mNG and ATTO647N-neurotensin in living *E. coli* cells** by structured illumination microscopy (SIM). **A, D,** *E. coli* expressing NTSR1-mNG in the presence of ATTO647N-labeled neurotensin, excited with 488 nm. **B, E,** same *E. coli* cells subsequently excited with 640 nm. **C, F,** overlay of images **A** and **B**, or **D** and **E**, respectively. Imaging of the two laser excitations for each z-stack with 11 planes. Image size is 36 x 36 µm$^2$.

### 3.3 Purification of NTSR1-mNG

The expression yield of NTSR1-mNG was relatively low compared to the expression of the thioredoxin fusion construct MBP-T43NTSR1-TrxA-H10 used in previous protocols [3, 16, 17]. However, an efficient purification strategy [4] allowed the isolation of pure NTSR1-mNG in quantities sufficient for the single-molecule studies. Due to the instability of the protein in detergent [3] all purification steps were executed at 4°C. Additionally, cholesteryl hemisuccinate was added to the solubilized protein to stabilize the neurotensin receptor [29].

The NTSR1-mNG purification is divided into four steps: Cell lysis and solubilization, nickel chelate affinity chromatography, neurotensin affinity chromatography and a final size exclusion chromatography step. In contrast to other membrane protein purifications such as purification of $F_OF_1$-ATP synthase [15, 30], cell lysis and membrane solubilization were done in one step. This strategy was rendered possible by the high efficiency of the subsequent purification step, i.e. nickel chelate affinity chromatography (Fig. 3: sub-panel "NiNTA" and Fig. 4 A). This step effectively captured the full-length receptor construct (Fig. 3: band 1) as well as a small fluorescent protein, which is most probably a degradation fragment comprising mNG (Fig. 3: band 3) with the decahistidine tag. Only one prominent non-fluorescent impurity was present in the NiNTA column eluate (Fig. 3: band 2).

The eluate from the nickel chelate affinity chromatography step was further purified *via* a custom-made NTS affinity column [3, 5] (Fig. 3: "NTS", Fig. 4 B). Binding affinities of neurotensin to the receptor are strongly modulated by Na$^+$ ions ranging from nanomolar affinities in the absence of Na$^+$ ions to no binding at 1 M Na$^+$ concentrations [31, 32]. Thus, NTSR1 binding is accomplished in buffer at a low Na$^+$ ion concentration, and elution at a high Na$^+$ ion concentration. The NTS column effectively separates the strong non-fluorescent impurity from NTSR1-mNG (Fig. 3: band 2) as well as the small degradation products (Fig. 3: band 3).

The final size exclusion chromatography step (Fig. 3 "SEC" and Fig. 4 C) resulted in a monodisperse peak indicating the high purity of the preparation.

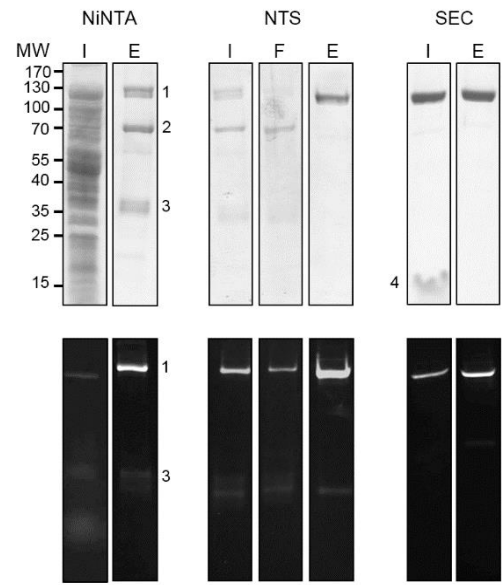

**Figure 3.** Purification of NTSR1-mNG. Samples of the major purification steps were subjected to 12 % SDS-PAGE. Before staining with coomassie (top), a fluorogram of the in-gel fluorescence of mNeonGreen was recorded (bottom). After cell lysis and solubilization of membranes, the solution was loaded on the NiNTA column (lane "I"). The NiNTA eluate (lane "E") shows the full-length construct (1) as well as lower molecular weight contaminants (3) and a non-fluorescent impurity (2). The NiNTA eluate was diluted and loaded on the NTS column (lane "I"). The impurity, the degradation product and a part of the full-length receptor did not bind (lane "F"). The NTS column eluate contained almost pure NTSR1-mNG (lane "E"). The final size exclusion column (SEC, lanes "I" and "E") removed a cloudy white low molecular weight band (4). On the left side, apparent molecular weights (MW) in kDa are indicated.

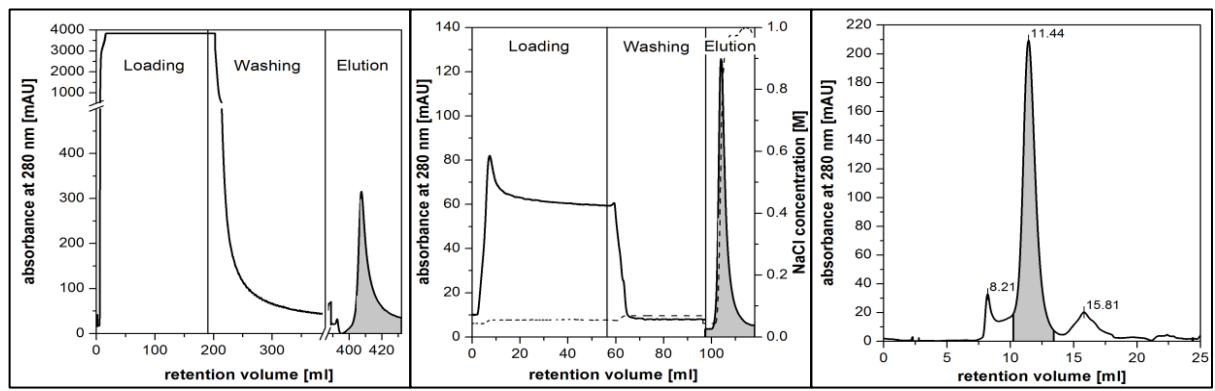

**Figure 4.** Chromatography steps of NTSR1-mNG purification. (**A**) Solubilized membrane proteins were loaded on a NiNTA sepharose column and washed extensively. The elution fractions (grey) containing the receptor fusion were pooled, diluted and loaded on a neurotensin column (**B**). After extensive washing, the receptors were eluted using high salt conditions [33]. The eluate fractions were pooled (grey), concentrated and loaded on a Superdex 200 10/300 GL column (**C**). The receptor preparation showed a monodisperse peak at 11.44 ml indicating its high purity. The small peak at the void volume (at 8.21 ml) probably contained high molecular weight aggregates and the peak at 15.81 ml contained a low molecular weight impurity which behaved like lipid or detergent on the SDS-PAGE. The area in grey was pooled and further concentrated.

### 3.4 Reconstitution of NTSR1-mNG into lipid vesicles

In order to reveal the binding of ATTO647N-NTS on the single molecule level by FRET, NTSR1-mNG was reconstituted into preformed liposomes at receptor concentrations of 10 nM. We used *E. coli* polar lipid extract for the liposomes instead of brain polar lipid extract because brain lipids contain a large fraction of unknown lipids that eventually interfere with single-molecule fluorescence measurements.

### 3.5 Binding of fluorescent NTS to single reconstituted NTSR1 in liposomes by smFRET

Single-molecule fluorescence recordings on a confocal microscope with a femtoliter-sized detection volume required a dilution of NTSR1-mNG (reconstituted at 10 nM) to about 100 pM. When fluorescent proteoliposomes traversed the detection volume a burst of photons was recorded. Shown in Fig. 5 A is a section of a time trace with several receptors observed in the presence of ATTO647N-labeled NTS. Individual photon bursts are marked with red numbers. Fluorescence of the FRET donor mNG ($I_D$) is plotted as a blue trace and that of the FRET acceptor ATTO647N ($I_A$) as a green trace. The proximity factor P is calculated for each data point with 1 ms length in the upper blue using the following equation (1):

$$P = (I_A) / (I_D + I_A) \qquad (1)$$

We subtracted a background on both traces for FRET donor and acceptor, but did not correct for instrument sensitivity nor different quantum yields nor spectral cross-talk to calculate the proximity factor P.

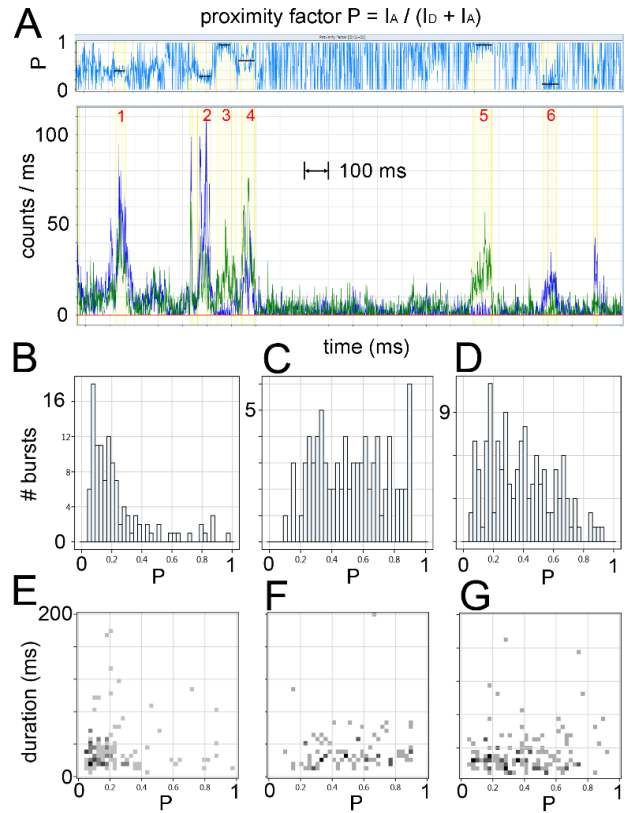

**Figure 5. A**, single-molecule FRET trace of individual NTSR1-mNG (blue intensity in lower panel) reconstituted into liposomes with bound ATTO647N-NTS (green intensity) freely diffusing in solution. Excitation with 491 nm (continuous wave); emission FRET donor (blue): 500-570 nm, FRET acceptor (green) for λ>595 nm, time bin 1 ms. The calculated proximity factor trace is shown in the upper panel (blue) with mean P (black line) for highlighted photon bursts. **B-D**, proximity factor histograms for NTSR1-mNG in the absence of ATTO647N-NTS (**B**), liposomes without NTSR1 in the presence of ATTO647N-NTS (**C**), and NTSR1-mNG with bound ATTO647N-NTS (**D**). **E-F**, 2D histograms of each photon burst duration in ms *versus* the proximity factor as shown in **B-D**.

In the presence of ATTO647N-NTS, photon bursts of reconstituted NTSR1-mNG with different P values were found (numbered in red in Fig. 5A). Burst #1 exhibited a nearly constant P~0.4, burst #2 a mean P~0.3, bursts #3 and #5 values of P>0.9, burst #4 an obviously fluctuating mean P~0.6, and burst #6 a P<0.1. Given an estimated Förster radius $R_0$~5 nm and fluorescence quantum yields ϕ for the FRET pair mNG (ϕ~0.85) and ATTO647N (ϕ~0.65), a negligible spectral cross-talk and a comparable detection efficiency for both markers was observed in our setup, and we attributed proximity factors 0.3<P<0.5 to distances between 4 and 5 nm. These are in the range of the expected distance for the ATTO647N fluorophore at the N-terminus of NTS on the extracellular side of NTSR1 and the chromophore of mNG on the intracellular side of the membrane.

We measured the distribution of proximity factors for reconstituted NTSR1-mNG in the absence of ATTO647N-NTS (Fig. 5 B) and for liposomes without the receptor (Fig. 5 C) as controls. Compared to the distribution of NTSR1-mNG with ATTO647N-NTS (Fig. 5 D), the histogram of NTSR1-mNG without ATTO647N-NTS showed a maximum of P values for P<0.2. Therefore, the related P values found in Fig. 5 D can be attributed to the "donor only" species, i.e. NTSR1-mNG without bound NTS. *Vice versa*, all bursts with P>0.3 are likely caused by FRET between ATTO647N-NTS and NTSR1-mNG. We noted that some liposomes made from *E. coli* polar lipids also exhibited photon bursts with apparent P values ranging across the entire spectrum between 0 and 1. Plotting the proximity factor of each photon burst *versus* the burst duration showed that similar sizes of the fluorescent particles were recorded, i.e. liposomes with or without NTSR1 exhibited burst durations between 20 and 40 ms (Fig. 5 E - G). Therefore, other fluorescence parameters, i.e. fluorescence lifetime using a pulsed laser, have to be recorded to identify and discriminate photon bursts showing apparent FRET in liposomes with and without NTSR1-mNG (as shown in chapter 3.8 below). Despite the small smFRET data set in Fig. 5 D, the broad distribution of FRET efficiencies may indicate a heterogeneous distribution of mNG orientations at the C-terminus of the receptor. An alternative explanation could be the existence of NTSR1-mNG dimers resulting in two FRET donors with different distances to a single bound ATTO647N-NTS. To evaluate this possibility, we determined the monomeric or oligomeric state of NTSR1-mNG.

### 3.6 Evaluating the monomeric and oligomeric state of NTSR1-mNG in detergent micelles

Previous FRET measurements in cuvettes using C-terminal fused enhanced cyan fluorescent protein as FRET donor or enhanced yellow fluorescent protein as FRET acceptor on NTSR1 indicated a constitutive dimer of NTSR1 in a membrane environment but a monomer in detergent solution [12]. FRET detection is also possible between two identical fluorophores, i.e. homoFRET. The measurement method of homoFRET is either time-resolved or static anisotropy that is suitable for single molecule analysis. Therefore, we investigated our NTSR1-mNG preparation by anisotropy measurements of single receptors in detergent solution.

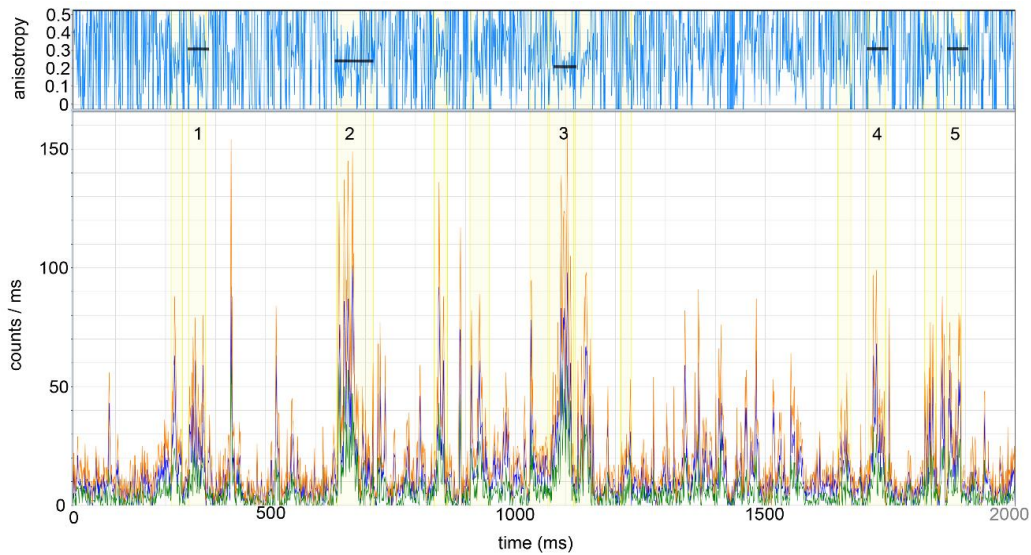

**Figure 6.** Anisotropy measurement of individual NTSR1-mNG in detergent solution. Linearly polarized pulsed excitation with 488 nm, 80 MHz, 50 µW. Lower panel shows parallel polarized fluorescence $I_{para}$ in blue, perpendicular polarized fluorescence $I_{perp}$ in green and sum intensity in orange. Upper panel shows the anisotropy trace. Time bin is 1 ms.

A time trace of a confocal anisotropy recording using freely diffusing 100 pM NTSR1-mNG in Tris/HCl buffer containing 1 mM EDTA and the detergents DDM (0.1%), CHAPS (0.5%) and CHS (0.1%) is shown in Fig. 6. Fluorescence is split into parallel ($I_{para}$, blue trace) and perpendicular polarization ($I_{perp}$, green trace). The sum of both is plotted as the orange trace. Single NTSR1-mNG exhibited peak intensities between 80 and 160 counts per ms in this selection. The upper blue trace is the anisotropy calculated per time bin of 1 ms. Anisotropy r is defined as (equation 2):

$$r = (I_{para} - I_{perp}) / (I_{para} + 2*I_{perp}) \quad (2)$$

For each photon burst defined by intensity thresholds, mean intensity, fluorescence lifetime τ of $I_{para}$, burst duration and anisotropy r was calculated. For example, burst #1 in Fig. 6 showed a mean intensity of 46 counts per ms, τ=2.8 ns, duration of 34 ms and anisotropy r=0.31. Burst #2 showed a mean intensity of 57 counts per ms, τ=3.0 ns, duration of 73 ms and anisotropy r=0.24; burst #3 showed a mean intensity of 69 counts per ms, τ=3.2 ns, duration of 48 ms and anisotropy

r=0.22; burst #4 showed a mean intensity of 45 counts per ms, τ=2.9 ns, duration of 34 ms and anisotropy r=0.31, and burst #5 showed a mean intensity of 46 counts per ms, τ=2.8 ns, duration of 29 ms and anisotropy r=0.31. While lifetimes τ were similar, the anisotropy r varied significantly across individual receptors.

For fluorescent proteins like mNG a high anisotropy r>0.3 is expected in solution because this is caused by the rigid protein environment of the chromophore suppressing rotational flexibility. Fusion of mNG to NTSR1 should result in similar or even higher r values due to the larger size of the receptor fusion protein. In contrast, changes of the anisotropy to lower values of r<0.3 could be caused by the presence of detergent which is changing the refractive index of the solution. Therefore, we compared soluble mNG and a tandem dimer of mNeonGreen (mNG5mNG) in the same detergent solutions as NTSR1-mNG. Fig. 7 summarizes the single-molecule anisotropy, lifetime and brightness data.

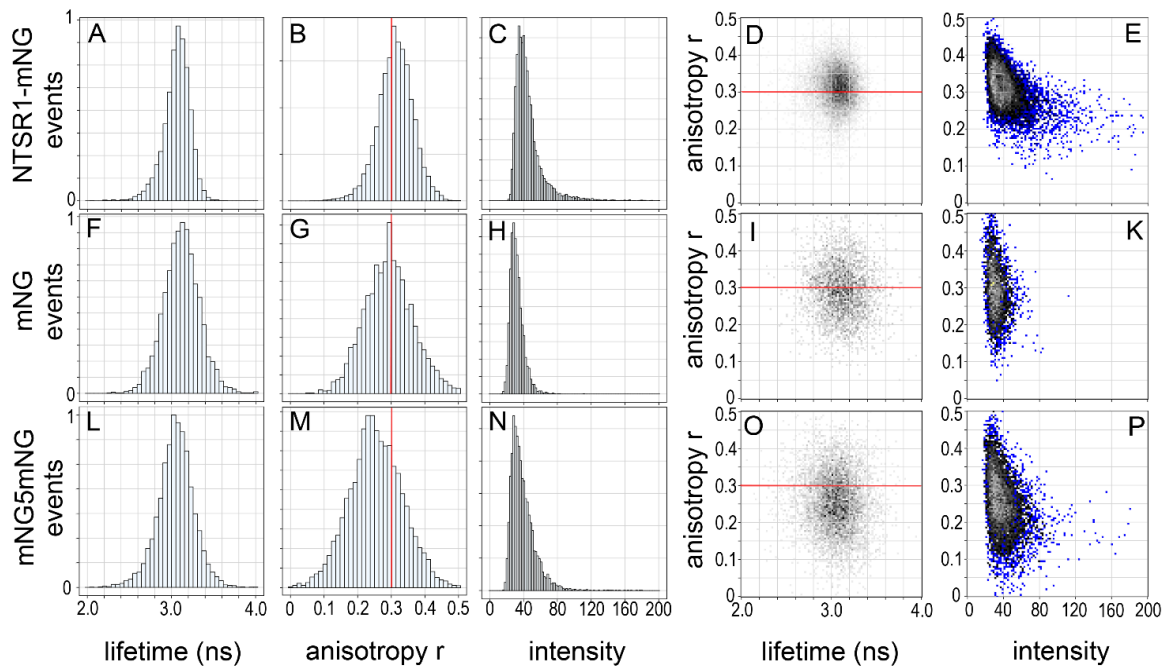

**Figure 7.** Normalized lifetime, anisotropy and intensity distributions for single NTSR1-mNG (**A-E**), single mNG (**F-K**) and individual dimeric mNG5mNG (**L-P**) in detergent solution. **A**, **F**, **L**, normalized lifetime τ distribution. **B**, **G**, **M**, normalized anisotropy r distribution; red line marks r=0.3. **C**, **H**, **N**, normalized distributions of mean sum intensity per burst (counts per ms). **D**, **I**, **O**, two-dimensional histograms of anisotropy versus lifetime; red line marks r=0.3. **E**, **K**, **P**, two-dimensional histograms of anisotropy versus mean intensity; blue spots mark single events. The frequency of events is normalized and coded in grey scale; it ranges from black (small numbers) to light grey (high numbers).

Fluorescence lifetimes were very similar for all three proteins and around τ~3.1 ns (Fig. 7 A, F, L). Anisotropies were distinct, with r~0.3 for the monomeric mNG and NTSR1-mNG, but r~0.22 for the dimeric mNG5mNG (Fig. 7 G, M). We noted that a fraction of mNG5mNG exhibited also higher anisotropies r>0.3 in detergent solution, probably due to the use of a highly flexible amino acid linker enabling variable relative orientations of the two chromophore transition dipole moments. The brightness of the dimer mNG5mNG was higher due to the increase in absorbance (i.e. twice the mNG monomer with intensities between 20 and 40 counts/ms). NTSR1-mNG showed a brightness distribution comparable to that of the mNG dimer, which was characterized by a population with mean intensities > 60 counts/ms in Fig. 7 C, N). However, the anisotropy distribution with a mean r>0.3 (Fig. 7 D) supported the notion that most receptors were monomeric in detergent micelles or did not show homoFRET. Importantly, the few very bright aggregates of individual NTSR1-mNG (shown as blue squares, with mean intensities > 80 counts/ms in Fig. 7 E) always exhibited anisotropies r<0.3 and thus could be identified by both high mean intensity as well as a correspondingly lowered anisotropy. We interpret the broad distribution of anisotropies for the monomeric mNG (Fig. 7 K) as likely caused by the low number of photons for each burst due to fast diffusion and the limited laser excitation power of 50 µW used (intended to avoid photobleaching and reduce blinking of mNG).

In summary, we compared the fluorescence properties of NTSR1-mNG to that of monomeric mNG and dimer mNG5mNG in detergent solution. Using pulsed laser excitation allowed to record fluorescence lifetime and time-resolved anisotropy in addition to intensities only. Single photon burst analysis revealed that NTSR1-mNG in detergent is mostly monomeric with anisotropies r>0.3. We note a fraction of oligomeric NTSR1-mNG with very high brightness, showing lower anisotropies and homoFRET, suggesting direct contact of the receptors.

### 3.7 Evaluating the monomeric and oligomeric state of NTSR1-mNG reconstituted in liposomes

We reconstituted NTSR1-mNG at low concentration (10 nM) in preformed liposomes in the presence of 200 mM NaCl (Fig. 8 lower panels). Because binding of NTS to NTSR1 is impeded at high NaCl concentrations, we also reconstituted NTSR1-mNG in the absence of NaCl (Fig. 8 upper panels). Both preparations showed a maximum of the anisotropy distribution around r=0.3 (Fig. 8 A, E). However, the reconstitution in the absence of NaCl resulted in a larger population of NTSR1-mNG with anisotropies lower than 0.3 possibly indicating a larger fraction of dimeric or oligomeric receptors. Comparison of the mean intensity distributions revealed that both preparations contained a significant number of NTSR1-mNG aggregates characterized by mean intensities larger than 80 counts/ms (Fig. 8 B, F). Differences were revealed by plotting lifetime versus anisotropy of single proteoliposomes (Fig. 8 C, G). Reconstitution in the absence of NaCl yielded a population of NTSR1-mNG with lower anisotropy and corresponding lower fluorescence lifetime τ~2.7 ns instead of τ~3.1 ns. For both reconstitution approaches the photon bursts with high mean counts exhibited an anisotropy r<0.3 (Fig. 8 D, H) indicating that NTSR1-mNG dimers and higher aggregates persisted during the reconstitution process and were not removed by the additional washing and ultracentrifugation steps applied to separate reconstituted from non-reconstituted receptors. As a consequence, any future single-molecule analysis of NTS binding has to be performed with improved liposome preparations containing monomeric NTSR1-mNG only, as assessed by brightness and anisotropy information.

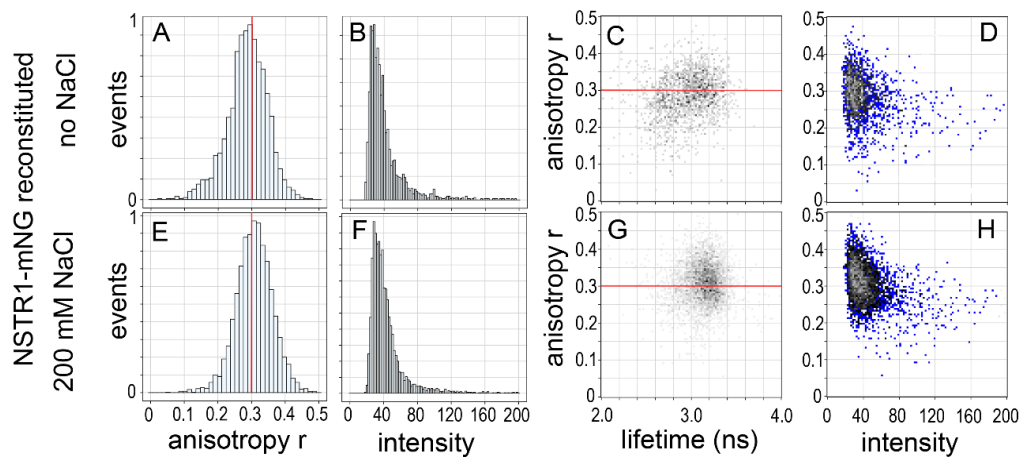

**Figure 8.** Normalized anisotropy, intensity and lifetime distributions for 10 nM NTSR1-mNG reconstituted in liposomes either in the absence of NaCl (**A-D**) or in the presence of 200 mM NaCl (**E-H**). **A, E**, normalized anisotropy r distribution; red line marks r=0.3. **B, F**, normalized distributions of mean sum intensity per burst (counts per ms). **C, G**, two-dimensional histograms of anisotropy versus lifetime; red line marks r=0.3. **D, H**, two-dimensional histograms of anisotropy versus mean intensity; blue spots mark single events. The frequency of events is normalized and coded in grey scale; it ranges from black (small numbers) to light grey (high numbers).

### 3.8 Measuring NTSR1-mNG by homoFRET and bound NTS by smFRET simultaneously

Finally, we combined the detection of monomeric and dimeric (or oligomeric) NTSR1-mNG species, present in a single liposome, with the detection of NTS by separating the FRET donor fluorescence of mNG into the parallel and perpendicular polarization components for anisotropy calculation (equation (2)) and simultaneously recording the FRET acceptor fluorescence with a third APD detector. Shown in Fig. 9 are four examples of photon bursts of NTSR1-mNG bound to ATTO594-NTS. ATTO594-NTS was used because of its higher fluorescence quantum yield compared to ATTO647N-NTS. The blue intensity trace is $I_{para}$ and the green trace is $I_{perp}$ of the FRET donor. The red trace is the FRET acceptor intensity. The upper blue trace is the proximity factor calculated by using the sum of both FRET donor channels.

The receptors from left to right exhibited anisotropy and donor lifetimes of r=0.25 and τ=3.3 ns with a proximity factor of P=0.30 for the photon burst on the left; the second burst had r=0.27 and τ=3.0 ns with P=0.48; the third burst had r=0.28 and τ=2.4 ns with P=0.42; and the burst on the right had r=0.36 and τ=1.9 ns with P=0.42. Photon burst intensities, anisotropies and lifetimes support the notion that NTSR1-mNG as FRET donor is monomeric for FRET to bound ATTO594-NTS. However, the total number of photons used to determine anisotropy and FRET donor lifetimes in a single photon burst were small resulting in large standard deviations. We note that very bright bursts were also present (data not shown), which we attributed to NTSR1-mNG aggregates. These latter data were not further analysed.

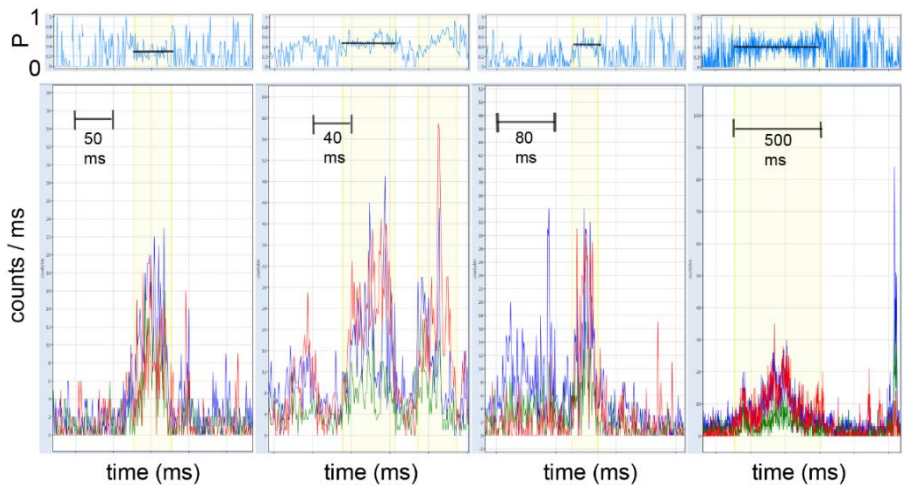

**Figure 9.** Examples of reconstituted NTSR1-mNG with bound ATTO594-NTS. For anisotropy measurements, the FRET donor intensity is split according to its polarization, i.e. in the $I_{para}$ (blue trace) and the $I_{perp}$ (green trace) components. FRET acceptor trace is plotted in red. The upper panels show the proximity factor traces calculated by FRET acceptor intensity divided by the sum of all three detection channels, and black lines mark the mean P value. Time bin is 1 ms.

## 4. DISCUSSION AND OUTLOOK

With the aim to study the ligand binding-induced conformational dynamics of the GPCR neurotensin receptor 1, we have constructed the vector pNTSR1_mNG that codes for the NTSR1-mNG fusion protein. Fusing tags to the C-terminus of NTSR1 does not affect the ligand binding properties of the receptor. To test ligand binding to NTSR1-mNG we applied structured illumination (SIM) for super-resolution microscopy to living *E. coli* cells. A truncated neurotensin fragment with a cysteine was labeled with a red fluorescent dye to yield ATTO647N-NTS, suitable for excitation by our SIM microscope laser. ATTO647N-NTS colocalized with membrane-bound NTSR1-mNG. However, the optical resolution of SIM in the range of 120 to 150 nm limits the interpretation of colocalization to these distances. Therefore, SIM experiments cannot discriminate between the specific binding of ATTO647N-NTS to NTSR1-mNG and non-specific retention of the fluorescent ligand.

FRET is the method of choice to detect colocalization within less than 8 nm. Confocal microscopy of freely diffusing proteoliposomes detected single-molecule FRET signals between NTSR1-mNG and ATTO647N-NTS (or ATTO594-NTS). However, a broad distribution of FRET efficiencies was found for individual receptors. As FRET efficiency is not only distance-dependent but also varies with the relative orientation of the transition dipole moments of the two fluorophores (i.e. $\kappa^2$) [34], the broad distribution of FRET efficiencies shown in Fig. 5 could be caused by the restricted mobility of the fluorophores. This is expected for mNG because of the large size of the receptor fusion protein and the fixed orientation of the chromophore within mNG. Furthermore, the N-terminally truncated NTS, used to generate ATTO647N-NTS, may restrict the rotational mobility of the attached ATTO647N dye when the peptide is bound within the receptor binding pocket, and $\kappa^2$ could be very different for each receptor. This effect is less likely using the full-length ligand ATTO594-NTS.

Alternatively, the reconstitution process using 10 nM NTSR1-mNG could have yielded a fraction of receptor dimers or multiple NTSR1-mNG monomers, inserted in a single liposome in opposite orientation. The receptors with the ligand binding pocket facing inside the liposome would thus not bind externally added NTS. In such a scenario, excess FRET donor fluorophores (NTSR1-mNG) over acceptor fluorophores (ATTO-NTS) would diminish the FRET efficiency, i.e.

resulting in lower proximity factors than the expected values of P~0.5. We evaluated the monomeric, dimeric and oligomeric states of NTSR1-mNG using single-particle homoFRET in detergent solution and after reconstitution in liposomes. As controls for homoFRET, we constructed and purified the soluble mNG and the dimeric mNG5mNG connecting two mNG proteins with a short five amino acid linker. Burst-wise analysis of confocal time-resolved fluorescence anisotropy data from mNG and mNG5mNG in the same detergent solution as NTSR1-mNG showed that all three proteins exhibited the same fluorescence lifetime, but different brightness and distinct fluorescence anisotropies. While the brightness distribution of NTSR1-mNG pointed towards a dimeric state of the receptor, the high anisotropy values of NTSR1-mNG did not result in homoFRET indicating no direct mNG contacts. This may result from several receptors inserted in a liposome, oriented up-side-down or separated by more than 10 nm in the lipid bilayer. Alternatively, we attribute the higher brightness of the receptor compared to soluble mNG to the slower diffusion and, accordingly, a longer mean transit time through the confocal detection volume. This was confirmed by a diffusion time comparison using FCS of the same single molecule measurements (data not shown). As expected from the apparent molecular weight of NTSR1-mNG detergent complex, diffusion times of the receptor were threefold longer than those of soluble mNG. Note that diffusion times scale with the radius of the particle, i.e. a tenfold difference in relative mass corresponds to an approximate threefold difference in diffusion time.

Many very bright aggregates of NTSR1-mNG in detergent solution were observed with low anisotropies r<<0.3, i.e. exhibiting homoFRET. After reconstitution of 10 nM NTSR1-mNG into liposomes the brightness distribution of individual proteoliposomes did not change with respect to the measurements in detergent (Fig. 7 B, F and 6 C) indicating that the populations of mostly monomeric receptors *versus* dimeric and oligomeric NTSR1-mNG were not altered. Aggregates with high brightness had low anisotropy, i.e. exhibited homoFRET for r<<0.3. Thus, we combined the detection of the monomeric state of reconstituted NTSR1-mNG by anisotropy and brightness measurements with the detection of bound full-length neurotensin ATTO594-NTS by smFRET using a modified confocal microscope setup with three APDs [22, 35]. Our preliminary results indicated that taking all fluorescence parameters (i.e. spectral information, brightness, lifetime, anisotropy) of individual receptors into account allows to reject those NTSR1-mNG with an equivocal dimeric or oligomeric state from subsequent single-molecule FRET analysis.

However, observation times of the proteoliposomes were short (i.e. 40 ms to a few hundred ms) and therefore yielded only a limited number of photons for the calculation of anisotropy for homoFRET or of lifetimes. In addition, the brightness of individual fluorophores cannot be measured quantitatively for freely diffusing particles as the intensity depends on the actual position within the laser focus or confocal detection volume, respectively. Constant intensities can only be achieved if the fluorescent molecule is not changing its relative position. We will apply two different approaches to obtain constant intensities and to prolong the observation times as well, either reconstituting NTSR1-mNG into free-standing lipid bilayers in a black-lipid-membrane setup [36], or using a trap for single molecules in solution that had been developed by A. E. Cohen and W. E. Moerner. The 'anti-Brownian electrokinetic trap' (ABELtrap) [37-42] combines a fast particle localization with a fast electrical feedback to electrodes in a microfluidic chip and allows to study single fluorescent molecules, proteins, proteoliposomes and more in solution for many seconds. The ABELtrap can record all fluorescence parameters of single particles with high photon numbers and yields their diffusion coefficient and charge [43-45]. Appropriate labeling of the purified NTSR1, of its ligand as well as the G protein, are prerequisites to unravel the conformational dynamics of this GPCR *in vitro* by recording multiparameter fluorescence of individual receptors one after another.

**Acknowledgements**

We thank I. Starke and H. Sielaff (Jena) for their help with image analysis of NTSR1-mNG in *E. coli*. The authors gratefully acknowledge financial support by the Deutsche Forschungsgemeinschaft DFG in the Collaborative Research Center/Transregio 166 "ReceptorLight" (to M.B.). The research of R.G. is supported by the Intramural Research Program of the National Institutes of Health, National Cancer Institute.